\documentclass{article}
\usepackage{spconf,amsmath,graphicx,hyperref}
\usepackage{amssymb}   
\usepackage{booktabs}   
\usepackage{siunitx}    
\usepackage{multirow}   
\usepackage{graphicx}   

\title{FAC-FACodec: Controllable Zero-Shot Foreign Accent Conversion with Factorized Speech Codec}
\name{Yurii Halychanskyi, Cameron Churchwell, Yutong Wen, Volodymyr Kindratenko}
\address{The Grainger College of Engineering, Siebel School of Computing and Data Science,\\
University of Illinois Urbana-Champaign, Urbana, IL, USA \\
\{yuriih2, cc178, yutong12, kindrtnk\}@illinois.edu}
\begin{document}
\ninept
\maketitle
\begingroup
\renewcommand\thefootnote{}
\footnotetext{Audio samples and additional materials are available at:
\url{https://claussss.github.io/accent_control_demo/}}
\addtocounter{footnote}{-1}
\endgroup
\begin{abstract}
Previous accent conversion (AC) methods, including foreign accent conversion (FAC), lack explicit control over the degree of modification. Because accent modification can alter the perceived speaker identity, balancing conversion strength and identity preservation is crucial. We present an AC framework that provides an explicit, user‑controllable parameter to adjust the strength of pronunciation-level accent modification. Results show performance comparable to recent AC systems, stronger preservation of speaker identity, and unique support for controllable accent conversion.
\end{abstract}

\section{Introduction}
\label{sec:intro}

Accents, together with timbre and speaking style, form an essential part of speaker identity~\cite{Lavan2019FlexibleVoices,Trudgill2006}. 
However, they can also create challenges in cross-regional communication, and speech technologies trained primarily on standard American or British English often do not transfer well to accented speech~\cite{69}. 
Accent conversion (AC) aims to mitigate these issues by transforming a speaker’s accent toward a target accent while preserving identity~\cite{17}. 
Foreign accent conversion (FAC) focuses specifically on adapting non-native accents into native ones.

Because accent is part of identity, altering it inevitably affects how the speaker is perceived. Applications therefore require different balances between modification and preservation: language learning may prioritize fully native-like pronunciation~\cite{2}, while dubbing requires duration and rhythm preservation for synchronization~\cite{28, accent_norm}, and personal communication may benefit from adjustable control. Thus, controllability is essential.

We introduce a diffusion-based framework that adapts prior noise-controlled speech editing techniques~\cite{75,77} to provide explicit control over the degree of accent modification. During training, the model learns a native-pronunciation prior by iteratively denoising Factorized Speech Codec (FACodec) phonetic representations~\cite{50} extracted from native speech under an ordinary differential equation (ODE) formulation. At inference, non-native representations are first corrupted with noise and then denoised toward the native prior. By choosing the initial noise level, users can smoothly adjust the degree of conversion from non-native to native-like pronunciation.

\textbf{Contributions.} 
(1) A framework for converting arbitrary source accents to a fixed target accent at the pronunciation level using noise-controlled diffusion editing.
(2) A training and inference pipeline that requires only native audio with transcripts for training and non-native audio with transcripts for inference, without the need for parallel accented data. 
(3) To our knowledge, this is the first AC approach providing explicit user control over accent strength, enabling a tunable trade-off between accent conversion and speaker identity preservation.
         
\section{Background}
\label{sec:background}

\subsection{Noise-Controlled Translation with Diffusion Priors}
Diffusion priors enable task-adaptable translation by adding noise to inputs and denoising them toward a learned distribution. SDEdit~\cite{75} showed this for images, where partial noising of sketches followed by iterative denoising produced realistic outputs, with the initial noise level acting as a control knob between input faithfulness and realism. In speech, similar ideas appear in editing frameworks such as EdiTTS~\cite{77}, which perturb latent representations and denoise them for controlled pitch or content changes.

Accent conversion naturally fits this paradigm. Deja et al.~\cite{76} proposed a diffusion-based many-to-many AC system: accent-salient regions are masked, noised, and denoised conditioned on target-accent embeddings, trained with parallel accented data. In contrast, our work addresses any-to-one AC using only native speech with transcripts, and explicitly introduces user-controllable noise as a mechanism to modulate accent strength—allowing users to balance conversion against identity preservation.

\subsection{FACodec: Factorized Neural Speech Codec}
Recent speech codecs produce compact latents that disentangle different aspects of speech. FACodec~\cite{50} encodes waveforms into an intermediate representation \(h\), which is factorized into content (\(z_c\)), prosody (\(z_p\)), and acoustic details (\(z_d\)), plus a global timbre embedding \(g\). This separation enables targeted modeling of pronunciation without altering prosody or timbre.

The content latent is split into two residuals, \(z_c = z_{c1}+z_{c2}\). In our framework, we operate on the quantized 8-dimensional representation of \(z_{c1}\), while \(z_{c2}\) is later regenerated from the denoised \(z_{c1}\). All other latents (\(z_p, z_d, g\)) remain fixed, allowing accent modification to focus solely on pronunciation.

\begin{figure}[t]
  \centering
  \includegraphics[width=0.88\columnwidth,trim=6pt 8pt 6pt 4pt,clip]{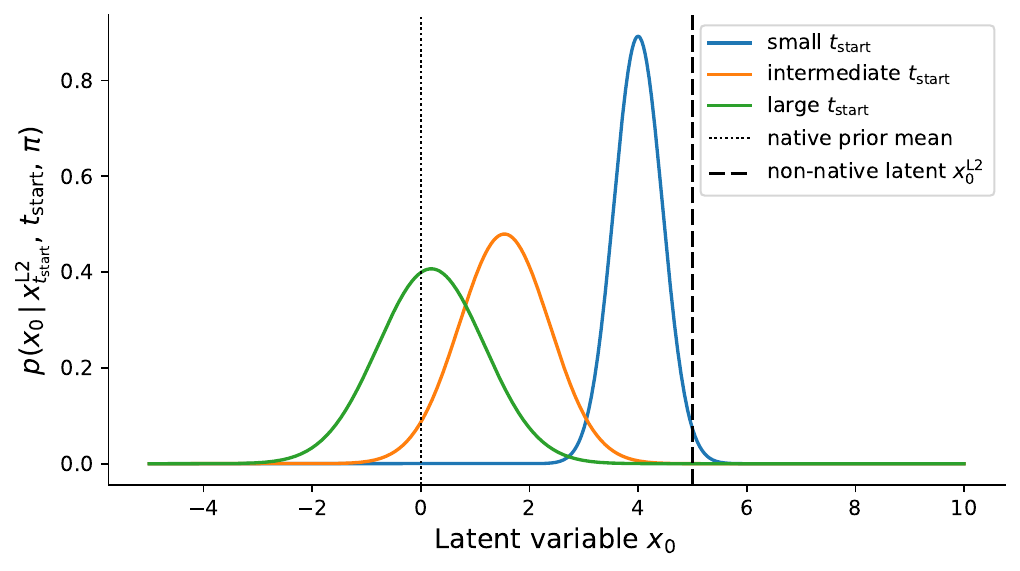}
  \caption{%
  Illustration of how $t_{\text{start}}$ controls a single-frame posterior.
  Larger $t_{\text{start}}$ widens the likelihood and shifts the posterior toward the native prior mean, while smaller $t_{\text{start}}$ keeps it concentrated near the non-native input.}
  \label{fig:posterior_plot}
  \vspace{-2mm}
\end{figure}

\section{Proposed Framework}
\label{sec:method}

\subsection{Forward Diffusion Process}
We operate on the FACodec content residual \(z_{c1}\), denoted \(x_0\); its noisy counterpart at timestep \(t\) is \(x_t\).
Native-accented training pairs are sampled from the empirical joint distribution:
\[
(x_0^{(n)}, \pi^{(n)}) \sim p_{\text{nat}}(x_0,\pi), \quad n=1,\dots,N,
\]
where \(\pi\) denotes aligned phoneme embeddings.
With a discrete schedule \(\{\beta_t\}_{t=0}^{T-1}\), increasing with \(t\) (where \(\alpha_t{=}1{-}\beta_t\) decreases accordingly and \(\bar\alpha_t{=}\prod_{s=0}^{t}\alpha_s\)), we form noisy latents
\begin{equation}
    x_t \;=\; \sqrt{\bar{\alpha}_t}\,x_0 \;+\; \sqrt{1-\bar{\alpha}_t}\,\varepsilon,\quad \varepsilon \sim \mathcal{N}(0,I).
    \label{eq:forward}
\end{equation}
Marginalizing over \(x_0\) yields the native-smoothed marginal at timestep \(t\):

\begin{equation}
\begin{aligned}
  p_t^{\text{nat}}(x_t \mid \pi) 
  &= \int p_{\text{nat}}(x_0 \mid \pi)\,
     \mathcal{N}\!\left(x_t \mid \sqrt{\bar{\alpha}_t}\,x_0,
     (1-\bar{\alpha}_t)I\right)\,dx_0.
\end{aligned}
\label{eq:marginal}
\end{equation}

\subsection{Conditional Diffusion Model}
Given \(x_t\), \(t\), and \(\pi\), a denoiser network \(s_\theta\) predicts the noise:
\[
s_\theta(x_t,t,\pi)\;\approx\;\mathbb{E}[\varepsilon \mid x_t,t,\pi],
\]
optimized with
\begin{equation}
    \mathcal{L}_{\text{diff}} \;=\; \mathbb{E}_{t,x_0,\varepsilon}\!\left[\left\|\varepsilon - s_\theta(x_t,t,\pi)\right\|_2^2\right].
\end{equation}

\subsection{Posterior Derivation}
Applying Bayes’ rule under phoneme conditioning gives
\begin{align}
    p(x_0\mid x_t,t,\pi) 
    &= \frac{p_{\text{nat}}(x_0\mid\pi)\,\mathcal{N}\!\left(x_t\mid \sqrt{\bar{\alpha}_t}\,x_0,(1-\bar{\alpha}_t)I\right)}{p_t^{\text{nat}}(x_t\mid\pi)}
    \label{eq:bayes_full}\\[-1pt]
    &\propto \underbrace{p_{\text{nat}}(x_0\mid\pi)}_{\text{native prior}}\;
    \underbrace{\exp\!\Big(-\tfrac{\|x_t-\sqrt{\bar{\alpha}_t}\,x_0\|_2^2}{2(1-\bar{\alpha}_t)}\Big)}_{\text{likelihood}}.
    \label{eq:bayes_simplified}
\end{align}
Since the denominator in \eqref{eq:bayes_full} does not depend on \(x_0\), we work with the proportional form \eqref{eq:bayes_simplified}.

\subsection{Inference Procedure and Accent-Strength Control}
Given a non-native sequence \(x_0^{\mathrm{L2}}\), we pick an initial timestep \(t_{\text{start}}\), corrupt via \eqref{eq:forward} to obtain \(x_{t_{\text{start}}}^{\mathrm{L2}}\), and run deterministic Denoising Diffusion Implicit Models (DDIM) with the ODE formulation~\cite{ddim} for \(K\) steps. At step \(t\),
\begin{equation}
    \hat{x}_0(x_t,t,\pi) \;=\; \frac{x_t - \sqrt{\,1-\bar{\alpha}_t\,}\,s_\theta(x_t,t,\pi)}{\sqrt{\bar{\alpha}_t}},
    \label{eq:x0_reconstruct}
\end{equation}
and update
\begin{equation}
    x_{t-1} \;=\; \sqrt{\bar{\alpha}_{t-1}}\,\hat{x}_0(x_t,t,\pi) \;+\; \sqrt{\,1-\bar{\alpha}_{t-1}\,}\,s_\theta(x_t,t,\pi).
    \label{eq:ddim_update}
\end{equation}

The control \(t_{\text{start}}\) balances likelihood and prior in \eqref{eq:bayes_simplified}: small values (narrow likelihoods) anchor solutions near the non-native input \(x_0^{\mathrm{L2}}\), while larger values broaden the likelihood and bias the posterior toward the native prior. This follows from the forward process, where the likelihood variance is \((1-\bar{\alpha}_{t_{\text{start}}})I\): smaller \(t_{\text{start}}\) yields smaller variance, larger \(t_{\text{start}}\) yields larger variance, as shown in Fig.~\ref{fig:posterior_plot}.

Each latent frame (20\,ms of audio) has its own posterior and is sampled independently. Even if the posterior mean shifts toward the prior at high \(t_{\text{start}}\), random draws still sometimes fall near the likelihood (non-native region). With hundreds of frames per utterance, this variability ensures that some segments remain more native-like while others retain traces of the original accent. Across the many independent frame-level samples in an utterance, this variability produces a perceptually smooth gradient of accent strength as \(t_{\text{start}}\) increases.

\subsection{Second Residual Prediction and Training Objective}
To improve \(z_c\) fidelity, we also predict \(z_{c2}\) from the denoised \(\hat{z}_{c1}\) (i.e., \(x_0\)) and shared encoder features \(h\):
\[
\hat{z}_{c2} = q_\phi\!\left(\text{concat}[\,h,\,\text{detach}(\hat{z}_{c1})\,]\right),
\]
with total loss
\[
\mathcal{L}_{\text{total}} = \mathbb{E}\!\left[\|\varepsilon - s_\theta(x_t,t,\pi)\|_2^2 + \lambda\,\|\hat{z}_{c2} - z_{c2}\|_2^2\right],\;\; \lambda=0.5.
\]
At inference, after \(K\) denoising steps yield \(\hat{z}_{c1}\), we predict \(\hat{z}_{c2}\) once, snap \(\hat{z}_{c1}\) to its nearest codebook vector, and form \(\hat{z}_c=\hat{z}_{c1}{+}\hat{z}_{c2}\). We keep \(z_p, z_d, g\) fixed and decode with FACodec to obtain the waveform. (Architecture and hyperparameters are detailed in Section~\ref{sec:architecture}.)

\section{Experiments}
\label{sec:experiments}

\subsection{Dataset}
\label{sec:dataset}
We train the denoising network on the LJSpeech corpus~\cite{64}, which contains 13,100 clips (about 24 hours) of a single female speaker with a Standard American English accent reading prepared text. This single-speaker design provides a consistent pronunciation target. We use 11,790 samples for training and 1,310 for validation.

Evaluation is performed on L2-Arctic~\cite{65}, which covers 24 speakers across 6 non-native accents (Hindi, Mandarin, Vietnamese, Korean, Spanish, Arabic). Each speaker reads 1,152 shared sentences, giving about 3 hours per accent.

Since source code is unavailable for prior AC baselines, we construct test sets from their demo pages. The sets for~\cite{30} and~\cite{23} contain subsets of L2-Arctic: 
\cite{30} includes 4 samples each for Arabic, Chinese, Korean, and Vietnamese, and 50 for Hindi; 
\cite{23} includes 4 samples for Chinese and Korean and 10 for Hindi. 
The set for~\cite{28} comes from three YouTube videos on its demo page, which we segmented into 45 clips (Chinese/Hindi) and 30 (Vietnamese).

\subsection{Model Architecture and Training Configuration}
\label{sec:architecture}
The denoiser is a 6-layer Transformer with 8 heads, model dimension 1024, and feed-forward network dimension 2048; dropout 0.1 is applied throughout. Phoneme conditioning is obtained by normalizing transcripts with Nvidia’s text normalization~\cite{61,66}, converting to phonemes with phonemizer+eSpeak-ng~\cite{62} (\texttt{en-us} accent), and aligning with Wav2Vec2 XLSR~\cite{XLSR}. Alignments are upsampled by nearest-neighbor interpolation to match FACodec frame resolution. Conditioning is injected via FiLM~\cite{58} and additive embeddings.

Content latents \(z_{c1},z_{c2}\) are standardized using dataset statistics. We adopt a linear noise schedule with \(\beta_t\in[10^{-4},2\!\times\!10^{-2}]\) over \(T{=}100\) steps. Inference uses the same ODE solver for \(K{=}100\) steps. Training uses Adam optimization algorithm with learning rate \(5\times10^{-5}\), batch size 64, for 360 epochs on a single Nvidia A40 GPU.

\subsection{Baseline Models}
We compare against three prior systems. The model in~\cite{30} modifies pronunciation only, while~\cite{28,23} also regenerate durations and intonation. The latter may sound more native but lose speaker style. None provide explicit control over accent strength, and all require accented data. We use the test sets described in Section~\ref{sec:dataset}.

\subsection{Evaluation Metrics}

We use both objective and subjective metrics. Content accuracy is measured with word error rate (WER) using Whisper~\cite{whisper} and JiWER.\footnote{\url{https://github.com/jitsi/jiwer}} Accent strength is estimated with an accent classifier~\cite{commonaccent,XLSR,speechbrain} trained on 16 accents, using the American English probability as a proxy for nativeness. Because the classifier is a black-box model and may exhibit dataset or accent biases, we use it only to analyze relative trends across conditions rather than to make absolute claims about accent quality. Speaker similarity is assessed with cosine similarity of WavLM x-vector embeddings~\cite{WavLM}. Unless otherwise noted, metrics are reported for five conditions: FACodec reconstruction and denoising with start timesteps \(t_{\text{start}}=25,50,75,100\). For fair comparison with prior work, we additionally report results at \(t_{\text{start}}=100\), which corresponds to the strongest accent modification and best matches baseline systems without a controllable knob.

For subjective evaluation, we conduct a MUSHRA-like test~\cite{mushra,reseval} with 13 participants. Each trial presented five versions of the same utterance (same content and speaker) corresponding to reconstruction and denoising with \(t_{\text{start}}=25,50,75,100\), which listeners rated simultaneously on a 0--100 scale for similarity to Standard American accent. Because no parallel native reference was available, no reference condition was provided; FACodec reconstruction (no conversion) served as a pseudo-anchor without additional degradation. Each participant completed 10 trials drawn at random from a pool of 340 utterances, yielding little overlap across listeners. Given the limited number of participants, these subjective results should be interpreted as indicative rather than conclusive. In the following tables, we abbreviate word error rate as WER, speaker similarity as SS, and accent classifier score as Acc, and we report $\Delta$, the change in accent score relative to the source input.

\section{Results}

\subsection{Controllability}
We first evaluate controllability on L2-Arctic across accent–conversion strengths \(t_{\text{start}}\) (Table~\ref{tab:objective}). The \emph{reconstruction} condition is FACodec encode–decode without conversion; because FACodec learns disentangled latents, reconstruction can bias pronunciation toward accents seen in codec training, yielding nonzero accent deltas. 

Averaged across accents, increasing \(t_{\text{start}}\) raises the American-accent classifier score while trading off with speaker similarity and WER. From reconstruction to \(t_{\text{start}}{=}100\), Acc increases from \(70.51\) to \(89.86\) (+19.35), SS decreases from \(0.98\) to \(0.88\) (–0.10), and WER rises from \(0.05\) to \(0.15\) (+0.10). These changes progress smoothly over \(t_{\text{start}}{=}25,50,75,100\), confirming effective fine-grained control. Accents with high initial American scores (e.g., Spanish, Korean) exhibit smaller or negative changes, whereas those with lower initial levels (e.g., Hindi, Vietnamese) gain more with larger \(t_{\text{start}}\). The simultaneous rise in accent scores and decline in speaker similarity reflects the expected trade-off between stronger conversion and identity preservation, while the increase in WER can be attributed to the greater amount of noise introduced at higher \(t_{\text{start}}\), which makes denoising more difficult and leaves residual artifacts.

\subsection{Comparison with Prior Systems}
We compare at \(t_{\text{start}}{=}100\) for fairness to baselines without a control knob, using the evaluation subsets described in Section~\ref{sec:dataset} (Table~\ref{tab:objective-h2h}).

\begin{table*}[!t]
\centering
\caption{Objective results on L2-Arctic. Left: FACodec reconstruction (no conversion). 
Right: denoising from different $t_{\text{start}}$ values (out of 100), with larger $t_{\text{start}}$ implying stronger conversion.}
\label{tab:objective}
\begingroup
\footnotesize
\sisetup{
    reset-text-series=false,
    text-series-to-math=true,
    mode=text,
    tight-spacing=true,
    round-mode=places,
    round-precision=2,
    table-number-alignment=center
}
\setlength{\tabcolsep}{2pt}
\renewcommand{\arraystretch}{0.92}
\newcommand{\accd}[2]{\num{#1}\,(\num{#2})}

\begin{tabular}{@{} 
l
S[table-format=1.2] S[table-format=1.2] c
S[table-format=1.2] S[table-format=1.2] c
S[table-format=1.2] S[table-format=1.2] c
S[table-format=1.2] S[table-format=1.2] c
S[table-format=1.2] S[table-format=1.2] c
@{}}
\toprule
& \multicolumn{3}{c}{\textbf{Reconstruction}} & \multicolumn{12}{c}{\textbf{$t_\text{start}$}} \\
\cmidrule(lr){2-4}\cmidrule(lr){5-16}
& \multicolumn{3}{c}{} & \multicolumn{3}{c}{\textbf{25}} & \multicolumn{3}{c}{\textbf{50}} & \multicolumn{3}{c}{\textbf{75}} & \multicolumn{3}{c}{\textbf{100}} \\
\cmidrule(lr){5-7}\cmidrule(lr){8-10}\cmidrule(lr){11-13}\cmidrule(lr){14-16}
\textbf{Accent}
& \textsf{WER} & \textsf{SS} & \textsf{Acc (\(\Delta\))}
& \textsf{WER} & \textsf{SS} & \textsf{Acc (\(\Delta\))}
& \textsf{WER} & \textsf{SS} & \textsf{Acc (\(\Delta\))}
& \textsf{WER} & \textsf{SS} & \textsf{Acc (\(\Delta\))}
& \textsf{WER} & \textsf{SS} & \textsf{Acc (\(\Delta\))} \\
\midrule
Arabic
& 0.04 & 0.98 & \accd{79.42}{-5.56}
& 0.06 & 0.97 & \accd{81.60}{-3.38}
& 0.07 & 0.95 & \accd{88.30}{ 3.31}
& 0.09 & 0.92 & \accd{93.27}{ 8.29}
& 0.13 & 0.90 & \accd{94.42}{ 9.44} \\
Chinese
& 0.07 & 0.98 & \accd{83.80}{ 2.14}
& 0.10 & 0.97 & \accd{84.61}{ 2.96}
& 0.12 & 0.95 & \accd{88.79}{ 7.13}
& 0.15 & 0.92 & \accd{93.63}{11.97}
& 0.19 & 0.91 & \accd{92.97}{11.31} \\
Hindi
& 0.03 & 0.99 & \accd{10.30}{ 4.28}
& 0.05 & 0.97 & \accd{16.16}{10.14}
& 0.06 & 0.92 & \accd{42.96}{36.94}
& 0.07 & 0.84 & \accd{70.20}{64.18}
& 0.11 & 0.79 & \accd{76.63}{70.61} \\
Korean
& 0.04 & 0.97 & \accd{97.42}{ -1.73}
& 0.06 & 0.95 & \accd{97.61}{ -1.53}
& 0.06 & 0.92 & \accd{96.57}{ -2.58}
& 0.07 & 0.90 & \accd{93.92}{ -5.22}
& 0.11 & 0.88 & \accd{90.64}{ -8.51} \\
Spanish
& 0.04 & 0.99 & \accd{91.16}{ -2.58}
& 0.05 & 0.97 & \accd{93.45}{ -0.28}
& 0.06 & 0.96 & \accd{94.77}{ 1.04}
& 0.08 & 0.93 & \accd{96.76}{ 3.02}
& 0.13 & 0.91 & \accd{95.93}{ 2.19} \\
Vietnamese
& 0.09 & 0.99 & \accd{60.97}{ 7.88}
& 0.12 & 0.97 & \accd{59.90}{ 6.81}
& 0.14 & 0.96 & \accd{73.61}{20.53}
& 0.16 & 0.94 & \accd{87.20}{34.11}
& 0.20 & 0.92 & \accd{88.58}{35.49} \\
\midrule   
Average
& 0.05 & 0.98 & \accd{70.51}{0.74}
& 0.07 & 0.97 & \accd{72.22}{2.45}
& 0.08 & 0.94 & \accd{80.83}{11.06}
& 0.10 & 0.91 & \accd{89.16}{19.39}
& 0.15 & 0.88 & \accd{89.86}{20.09} \\
\bottomrule
\end{tabular}

\endgroup
\end{table*}

\begin{table}[!t]  
\centering
\caption{Head-to-head comparison with prior systems on their respective small evaluation subsets derived from public demos. For each baseline, both the competitor system (\textsf{Comp.}) and our method (\textsf{Ours}) are evaluated on the same subset, with our method using $t_{\text{start}}{=}100$, to enable a fair within-method comparison. Results are not directly comparable across different baselines or with the full L2-Arctic evaluation in Table~\ref{tab:objective}.}
\label{tab:objective-h2h}
\begingroup
\footnotesize
\sisetup{
  reset-text-series=false,
  text-series-to-math=true,
  mode=text,
  tight-spacing=true,
  round-mode=places,
  round-precision=2
}
\setlength{\tabcolsep}{2pt}
\renewcommand{\arraystretch}{0.92}

\newcommand{\accd}[2]{\num{#1}\,(\num{#2})}

\resizebox{\columnwidth}{!}{%
\begin{tabular}{@{}  
  l
  l
  S[table-format=1.2] S[table-format=1.2]  
  S[table-format=1.2] S[table-format=1.2]  
  c c                                      
@{}}
\toprule
\textbf{Competitor} & \textbf{Accent}
& \multicolumn{2}{c}{\textbf{WER}}
& \multicolumn{2}{c}{\textbf{SS}}
& \multicolumn{2}{c}{\textbf{Acc\,(\(\Delta\))}} \\
\cmidrule(lr){3-4}\cmidrule(lr){5-6}\cmidrule(lr){7-8}
& & \textsf{Comp.} & \textsf{Ours}
  & \textsf{Comp.} & \textsf{Ours}
  & \textsf{Comp.} & \textsf{Ours} \\
\midrule
\multirow{3}{*}{\cite{28}}
  & Chinese    & 0.19 & 0.08 & 0.87 & 0.87 & \accd{98.66}{55.76} & \accd{69.68}{26.79} \\
  & Hindi      & 0.15 & 0.13 & 0.77 & 0.83 & \accd{99.87}{90.31} & \accd{72.41}{62.85} \\
  & Vietnamese & 0.19 & 0.11 & 0.86 & 0.88 & \accd{96.60}{51.92} & \accd{52.44}{7.76}  \\
  \midrule   
  & Average & 0.18 & 0.11 & 0.83 & 0.86 & \accd{98.38}{66.00} & \accd{64.84}{32.47}  \\
\midrule
\multirow{5}{*}{\cite{30}}
  & Arabic     & 0.14 & 0.15 & 0.86 & 0.82 & \accd{99.68}{53.13} & \accd{96.64}{50.09} \\
  & Chinese    & 0.02 & 0.00 & 0.82 & 0.93 & \accd{94.67}{23.08} & \accd{100.00}{28.41} \\
  & Hindi      & 0.06 & 0.09 & 0.62 & 0.79 & \accd{99.65}{66.58} & \accd{87.60}{54.53} \\
  & Korean     & 0.05 & 0.02 & 0.85 & 0.92 & \accd{49.72}{-50.26} & \accd{99.99}{0.01}  \\
  & Vietnamese & 0.21 & 0.08 & 0.84 & 0.95 & \accd{99.91}{15.02} & \accd{100.00}{15.11} \\
  \midrule   
  & Average & 0.10 & 0.07 & 0.80 & 0.88 & \accd{88.73}{21.51} & \accd{96.85}{29.63} \\
\midrule
\multirow{3}{*}{\cite{23}}
  & Chinese    & 0.17 & 0.12 & 0.81 & 0.92 & \accd{98.72}{98.24} & \accd{39.61}{39.14} \\
  & Hindi      & 0.08 & 0.04 & 0.76 & 0.79 & \accd{66.89}{66.67} & \accd{49.89}{49.67} \\
  & Korean     & 0.03 & 0.11 & 0.78 & 0.91 & \accd{73.88}{26.06} & \accd{50.03}{2.21}  \\
  \midrule   
  & Average & 0.09 & 0.09 & 0.78 & 0.87 & \accd{79.83}{63.66} & \accd{46.51}{30.34}  \\
\bottomrule
\end{tabular}%
} 

\endgroup
\end{table}

\begin{figure}[t]
  \centering
  \includegraphics[width=\columnwidth]{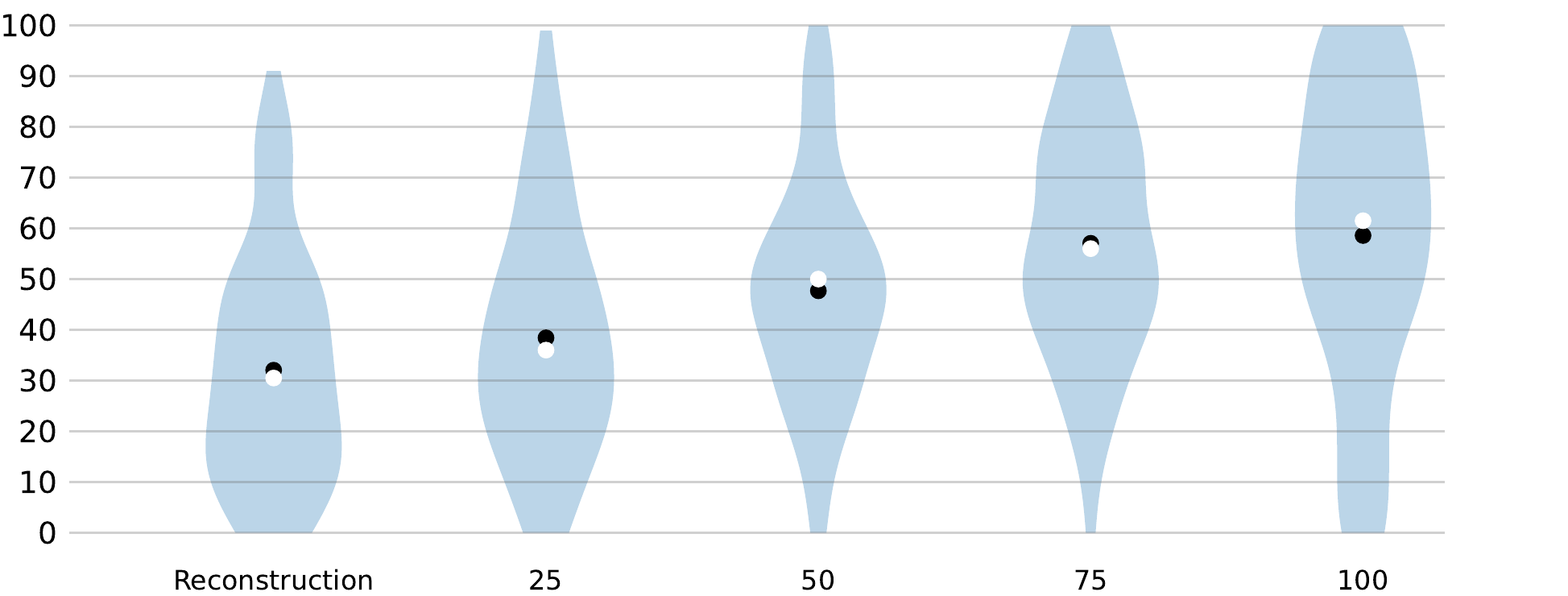}
    \caption{MUSHRA-like listening test results. Violin plots show subjective ratings of accent similarity to Standard American English for reconstruction and for denoising with $t_{\text{start}}=25,50,75,100$. Higher scores indicate stronger perceived similarity.}
  \label{fig:listening}
  \vspace{-2mm}
\end{figure}

\textbf{Pronunciation-only models.} The system proposed in~\cite{30} modifies pronunciation while retaining prosody. On the available subsets, our model shows higher speaker similarity, slightly lower word error rate, and comparable or higher American-accent confidence on average. For Korean, however, the competitor subset contains only four samples; in two of them the accent classifier labeled the output as \emph{Canadian} rather than \emph{American}, leading to a large negative accent–score \(\Delta\) even though perceptually the speech still sounded close to North American English. Aside from this, the results are broadly similar, but our method better preserves speaker identity while maintaining competitive intelligibility and accent conversion strength.

\textbf{Pronunciation and prosody regeneration.} The systems introduced in~\cite{28} and~\cite{23} regenerate native-like prosody in addition to modifying pronunciation. This design yields much higher American-accent classifier scores than ours on the same subsets, highlighting the contribution of prosody to perceived nativeness. However, our speaker-similarity metric is based on time-averaged timbre embeddings and does not reflect the loss of speaker-specific prosodic traits such as rhythm and intonation, which listeners do perceive. In practice, regenerated prosody discards the original style, whereas our approach leaves the original prosodic patterns unchanged by design. Objective results indicate that we match or surpass these systems in speaker similarity and maintain equal or lower word error rate, though at the cost of lower accent scores. Together, these outcomes illustrate the trade-off: regenerating prosody can increase perceived nativeness, while pronunciation-focused editing leaves the original prosodic structure unchanged, which may help preserve aspects of the speaker’s original style.

\subsection{Subjective Evaluation}
We also report results from the listening test described in Section~\ref{sec:dataset}. The results align with the objective metrics: listeners judged higher \(t_{\text{start}}\) settings as progressively more native-sounding, while reconstruction received the lowest scores. 
As shown in Fig.~\ref{fig:listening}, the distribution of ratings forms a perceptually smooth gradation of accent strength rather than abrupt changes, confirming that the control knob behaves as intended.

\section{Conclusion and Future Work}

We presented the first accent conversion framework with an explicit, user-controllable strength parameter. Trained only on native speech with transcripts, our diffusion ODE model operates on FACodec content latents and, at inference, applies controlled noise and iterative denoising from a chosen start timestep \(t_{\text{start}}\). This design yields perceptually smooth interpolation between the original non-native articulation and more native-like pronunciation. Objective metrics and listening tests confirm competitive performance and highlight the unique benefit of controllability.

Future directions include (i) \emph{perceptually guided noise}, where corruption is concentrated on strongly accented regions while leaving native-like segments minimally perturbed, potentially reducing reliance on transcripts, and (ii) \emph{coordinated suprasegmental control}, where prosody and acoustic-detail latents are regenerated jointly with the content latent to extend controllability beyond segmental pronunciation to suprasegmental aspects such as intonation and rhythm.

\section*{Acknowledgments}
This work used the Delta supercomputing system at the National Center for Supercomputing Applications (NCSA) through allocation from the ACCESS program, which is supported by National Science Foundation grants \#2138259, \#2138286, \#2138307, \#2137603, and \#2138296 \cite{access2023}.


\bibliographystyle{IEEEbib}
\bibliography{ref}

\end{document}